\documentclass[jkps,preprint,fleqn,showpacs,showkeysm,floatfix]{revtex4-1}
\usepackage{bm}
\usepackage{float}
\usepackage{amssymb,amsfonts,amsmath}
\begin{document}
\setcounter{page}{0}
\title[]{Entropy Production by Logarithmic Decomposition}
\author{Jang-il \surname{Sohn}}
\email{physicon@korea.ac.kr}
\affiliation{Department of Physics, Korea University, Seoul 136-713, Korea}

\date{\today}

\begin{abstract}
In statistical physics, entropy is generally logarithm of probability.
Therefore, if dynamics is decomposed by log, entropy production should be decomposed properly.
In the present work, log-decomposition of dynamics is introduced.
By which time evolution operator is logarithmically decomposed into a symmetric operator and an asymmetric factor.
Path probability and path entropy production are also systematically and intuitively decomposed into symmetric and asymmetric parts.
From symmetric operator, non-adiabatic entropy production is derived, whereas adiabatic entropy production is from asymmetric factor.
\end{abstract}



\maketitle

\section{Introduction and Purpose}

According to the 2'nd law of thermodynamics, average entropy change is always larger than or equal to zero, $\langle \Delta S \rangle \ge 0$, in macroscopic level.
However, because of fluctuation in microscopic level, there definitely exist entropy decreasing events which is explained very well by fluctuation theorems(FTs).
Moreover, the 2'nd law of thermodynamics can be derived from which \cite{PhysRevLett.104.090601,seifert2012stochastic}.
For the reasons, FTs are considered to be generalized versions of the 2'nd law of thermodynamics, so various types of FTs have been developed ever since the first FT was introduced by Evans, Cohen, and Morriss in 1993 \cite{PhysRevLett.71.2401}.


Most of FTs are based on the conjugated dynamics (so-called dual dynamics or time reversal transitions), 
\begin{equation}
\label{conjugated_dynamics}
w^{+}_{ij} \equiv (p^{st}_{i}/p^{st}_{j})w_{ji}
\end{equation}
where $w_{ji}$ is transition probability from $i$ to $j$, and $p^{st}_{i}$ is a probability distribution in a stationary state \cite{PhysRevLett.104.090601,seifert2012stochastic,ross1996stochastic}.
As a representative example, Esposito and Van den-Broeck \cite{PhysRevLett.104.090601} have defined total, adiabatic and non-adiabatic path entropy productions as $\Delta S_{tot} \equiv \ln(\mathcal{P}/\bar{\mathcal{P}})$, $\Delta S_{a} \equiv \ln(\mathcal{P}/\bar{\mathcal{P}}^{+})$ and $\Delta S_{na} \equiv \ln(\mathcal{P}/\mathcal{P}^{+})$, respectively.
The symbol ${}^{+}$ indicates the conjugated dynamics, and the symbol $\bar{}$ indicates reversed path.
From the definitions, three detailed FTs are derived,
\begin{eqnarray}
\label{detailed_FT}
\frac{P(\Delta S_{tot/na/a})}{P(-\Delta S_{tot/na/a})} &=& e^{\Delta S_{tot/na/a}} \\
\textrm{or}\quad \langle e^{\Delta S_{tot/na/a}} \rangle &=& 1
\end{eqnarray}
where $P(\Delta S)$ is probability of a event of $\Delta S$, and $\Delta S_{tot} = \Delta S_{na} + \Delta S_{a}$  \cite{PhysRevLett.104.090601}.
The non-adiabatic contribution consists of system entropy and excess entropy, $\Delta S_{na} = \Delta S_{sys} + \Delta S_{ex}$ \cite{PhysRevLett.86.3463,seifert2012stochastic,PhysRevE.60.R5017,oono1998steady}, and the adiabatic contribution is identical with house-keeping entropy, $\Delta S_{a} = \Delta S_{hk}$ \cite{oono1998steady,seifert2012stochastic,speck2005integral,chernyak2006path}.
Not only their FTs but also many other FTs are explained successfully by the conjugated dynamics, Eq. (\ref{conjugated_dynamics}).


However, in the present work, we propose another way which is logarithmic decomposition of dynamics.
We believe that proper FTs can be intuitively derived by logarithmic decomposition of dynamics, because entropy is defined by logarithm of probability.
That is reason we are performing this work.
By the log-decomposition of dynamics, path probability and path entropy production are also systematically decomposed.
The relevant FTs of log-decomposition are same with the three detailed or integral FTs.
In the present work, we use time-evolution operator (or propagator) instead of transition probability.


\section{Total Path Entropy Production}

We will calculate path entropy productions using time evolution operator instead of transition probability.
That makes our works simple, because there is no need to consider waiting time for calculating path probabilities.


Suppose that there is an ergodic system controlled by a schedule of a control parameter $\lambda$.
The system evolves along a path,
\begin{equation}
\label{path}
[\sigma]_0^T \equiv [i,\lambda]_{0}^{T} \dot =
\Big[ 
i_0 \xrightarrow {\lambda_{1}} i_{1} 
\xrightarrow {\lambda_{2}} \cdots 
\xrightarrow {\lambda_{T}} i_{T}
\Big],
\end{equation}
where $\sigma$ is a set of $i$ (state of the system) and $\lambda$.
The path probability in forward direction is given by
\begin{equation}
\label{path_probability}
\mathcal{P}[\sigma]_0^T = \prod_{t=1}^{T}w_{i_{t}i_{t-1}}(\lambda_{t}) p_{i_0}(0),
\end{equation}
where $w_{ij}(\lambda)$ is a time evolution operator under influence of $\lambda$.
If $i \neq j$, that plays a role of transition probability from $j$ to $i$.
However, if $ i = j$, that becomes a probability for staying at previous state $j$.
For a reversed path,
\begin{equation}
\label{rpath}
[\sigma]_T^0 \equiv [i,\lambda]_{T}^{0} \dot =
\Big[ i_0 \xleftarrow {\lambda_{1}} i_{1} 
\xleftarrow {\lambda_{2}} \cdots 
\xleftarrow {\lambda_{T}} i_{T} \Big],
\end{equation}
the reversed path probability is given by
\begin{equation}
\label{rpath_probability}
\mathcal{P}[\sigma]_T^0 = \prod_{t=1}^{T} w_{i_{t-1}i_{t}}(\lambda_{t}) p_{i_T}(T).
\end{equation}
In (\ref{rpath}) and (\ref{rpath_probability}), to indicate {\it reversed direction}, we use the notation $[\cdot]_{T}^{0}$ which means reversed path, $i.e.$ $\mathcal{P}[\sigma]_T^0 \equiv \bar {\mathcal{P}}[\sigma ]_0^T = {\mathcal{P}}[\bar \sigma ]_0^T$ where $\bar \sigma_{t} \equiv = ( i_{T-t},\lambda_{T-t})$.


Total path entropy production is defined by log of the forward and reversed path probabilities, (\ref{path_probability}) and (\ref{rpath_probability}), 
\begin{equation}
\label{total_path_entropy}
\Delta S_{tot}[\sigma]_0^T 
= \ln \frac{\mathcal{P}[\sigma]_0^T}{\mathcal{P}[\sigma]_T^0}.
\end{equation}


\section{Logarithmic Decomposition of Dynamics}

In the section, let us introduce log-decomposition of dynamics.
We decompose $w_{ij}(\lambda)$ into a symmetric operator, $\epsilon_{ij}$, and an log-asymmetric operator, $\nu_{ij}$.
The symmetric operator can be defined in any way, provided that detailed balance condition is satisfied in a steady state,
\begin{equation}
\label{eq:detailed_balance}
\epsilon_{ij}(\lambda) p^{st}_{j}(\lambda) = \epsilon_{ji}(\lambda) p^{st}_{i}(\lambda),
\end{equation}
where $p^{st}_{i}(\lambda)$ is a steady state distribution for fixed $\lambda$.
Since the symmetric operator satisfies detailed balance condition (\ref{eq:detailed_balance}), that can be considered to be an time evolution operator in an {\it effective equilibrium} \cite{zia2006possible,zia2007probability} or {\it effective Hamiltonian} \cite{hershfield1993reformulation,dutt2011effective,james2007effective,jolicard1995effective,PhysRevB.75.125122}.
Note that $\epsilon_{ij}(\lambda)$ is not actual equilibrium dynamics because $p^{st}_{i}(\lambda)$ is generally a nonequilibrium steady state distribution (or driven out of equilibrium).


Using the symmetric operator, we try log-decomposition of evolution operator such that
\begin{equation}
\label{eq:AS_operator}
\ln w_{ij}(\lambda) = \ln \epsilon_{ij}(\lambda) + \ln \nu_{ij}(\lambda).
\end{equation}
Here, we call $\nu_{ij}(\lambda)$ an asymmetric factor,
\begin{equation}
\label{def:nu}
\nu_{ij}(\lambda) \equiv \frac{w_{ij}(\lambda)}{\epsilon_{ij}(\lambda)}.
\end{equation}
If, and only if given system is in an (actual) equilibrium steady state, $\nu_{ij}(\lambda) = \nu_{ji}(\lambda)$.


By the log-decomposition of dynamics (\ref{eq:AS_operator}), forward path probability is also decomposed logarithmically,
\begin{equation}
\label{eq:AS_path}
\ln \mathcal{P}[\sigma]_0^T = \ln \mathcal{P}_{\epsilon}[\sigma]_0^T + \ln \mathcal{P}_{\nu}[\sigma]_0^T,
\end{equation}
where the path probability due to $\epsilon_{ij}(\lambda)$ is given by
\begin{equation}
\mathcal{P}_{\epsilon}[\sigma]_0^T \equiv \prod_{t=1}^{T} \epsilon_{i_{t}i_{t-1}}(\lambda_{t}) p_{i_0}(0),
\end{equation}
and a {\it path-probability-like} quantity is given by multiplication of the factor $\nu_{ij}(\lambda)$,
\begin{equation}
\mathcal{P}_{\nu}[\sigma]_0^T \equiv \prod_{t=1}^{T} \nu_{i_{t}i_{t-1}}(\lambda_{t}).
\end{equation}
Reversed path probability is decomposed in the same manner
\begin{equation}
\ln \mathcal{P}[\sigma]_T^0 = \ln \mathcal{P}_{\epsilon}[\sigma]_T^0 + \ln \mathcal{P}_{\nu}[\sigma]_T^0
\end{equation}
where
\begin{eqnarray}
\mathcal{P}_{\epsilon}[\sigma]_T^0 
&\equiv& \prod_{t=1}^{T} \epsilon_{i_{(T-t)}i_{(T-t+1)}}(\lambda_{(T-t+1)}) p_{i_T}(T) \nonumber \\
&=& \prod_{t=1}^{T} \epsilon_{i_{t-1}i_{t}}(\lambda_{t}) p_{i_T}(T)
\end{eqnarray}
and
\begin{eqnarray}
\mathcal{P}_{\nu}[\sigma]_T^0 
&\equiv& \prod_{t=1}^{T} \nu_{i_{(T-t)}i_{(T-t+1)}}(\lambda_{(T-t+1)}) \nonumber \\
&=&\prod_{t=1}^{T} \nu_{i_{t-1}i_{t}}(\lambda_{t}).
\end{eqnarray}


Therefore total entropy production (\ref{total_path_entropy}) is decomposed as follows
\begin{equation}
\label{eq:AS_entropy}
\Delta S_{tot}[\sigma]_0^T = \Delta S_{\epsilon}[\sigma]_0^T + \Delta S_{\nu}[\sigma]_0^T
\end{equation}
where the symmetric and asymmetric parts of path entropy production are
\begin{equation}
\label{symmetric_entropy}
\Delta S_{\epsilon}[\sigma]_0^T 
\equiv \ln \frac{\mathcal{P}_{\epsilon}[\sigma]_0^T}{\mathcal{P}_{\epsilon}[\sigma]_T^0}
=\ln \prod_{t=1}^{T} \frac{\epsilon_{i_{t}i_{t-1}}(\lambda_{t})p_{i_0}(0)}{\epsilon_{i_{t-1}i_{t}}(\lambda_{t})p_{i_T}(T)}
\end{equation}
and
\begin{equation}
\label{asymmetric_entropy}
\Delta S_{\nu}[\sigma]_0^T 
\equiv \ln \frac{\mathcal{P}_{\nu}[\sigma]_0^T}{\mathcal{P}_{\nu}[\sigma]_T^0}
= \ln \prod_{t=1}^{T} \frac{\nu_{i_{t}i_{t-1}}(\lambda_{t})}{\nu_{i_{t-1}i_{t}}(\lambda_{t})},
\end{equation}
respectively.



Just taking logarithm of time evolution operator, not only path probability but path entropy production are also decomposed simply, systematically and intuitively as seen in TABLE \ref{tab:AS_separation}.
Reminding that entropy is defined by logarithm of probability, log-decomposition makes sense, and indeed intuitive.

\begin{table}[h]
\caption{Logarithmic Separations} 
\centering 
\begin{tabular}{c c} 
\hline\hline 
time evolution operator & $\ln w_{ij} = \ln \epsilon_{ij} + \ln \nu_{ij}$ \\
path probability        & $\ln \mathcal{P} =\ln \mathcal{P}_{\epsilon} + \ln \mathcal{P}_{\nu}$ \\
path entropy production & $\Delta S_{tot} = \Delta S_{\epsilon} + \Delta S_{\nu}$ \\
\hline 
\end{tabular}
\label{tab:AS_separation}
\end{table}


\section{Three Integral or Detailed Fluctuation Theorems}
\label{three_FTs}

The FTs relevant to log-decomposition is same with three integral or detailed FTs.
Using (\ref{eq:detailed_balance}) and (\ref{def:nu}), it can be simply shown that the symmetric entropy production is identical with non-adiabatic contribution,
\begin{equation}
\label{eq:S_epsilon}
\Delta S_{\epsilon}[\sigma]_0^T = \ln \frac{\mathcal{P}_{\epsilon}[\sigma]_0^T}{\mathcal{P}_{\epsilon}[\sigma]_T^0} = \ln \frac{\mathcal{P}}{\bar{\mathcal{P}}^{+}}
= \Delta S_{na},
\end{equation}
and the asymmetric entropy production is adiabatic one,
\begin{equation}
\label{eq:S_nu}
\Delta S_{\nu}[\sigma]_0^T = \ln \frac{\mathcal{P}_{\nu}[\sigma]_0^T}{\mathcal{P}_{\nu}[\sigma]_T^0} = \ln \frac{\mathcal{P}}{\mathcal{P}^{+}}= \Delta S_{a}.
\end{equation}
Therefore the relevant FTs also naturally hold,
\begin{eqnarray}
\frac{P(\Delta S_{tot/\epsilon/\nu})}{P(-\Delta S_{tot/\epsilon/\nu})} &=& e^{\Delta S_{tot/\epsilon/\nu}} \\
\textrm{or}\quad 
\label{integral_log_entropy}
\langle e^{\Delta S_{tot/\epsilon/\nu}} \rangle &=& 1.
\end{eqnarray}
Applying Jensen's inequality \cite{ross1996stochastic} to Eq. (\ref{integral_log_entropy}), the 2'nd law of thermodynamics is derived, 
$$\langle \Delta S_{tot/\epsilon/\nu} \rangle \ge 0$$
where each average entropy change is given by path integrations over all possible paths, $$\langle \Delta S_{tot/\epsilon/\nu} \rangle = \sum_{[\sigma]} \mathcal{P}[\sigma]_{0}^{T} \Delta S_{tot/\epsilon/\nu} [\sigma]_{0}^{T}.$$


\section{A Simple Example: Entropy Production in Asymmetric Random Walk}

Generally, $\Delta S_{na}$ is defined as
\begin{equation}
\Delta S_{na} = \ln \frac{p_{i_0}(0)}{p_{i_T}(T)} + \ln \prod_{t=1}^{T} \frac{p^{st}_{i_{t}}(\lambda_{t})}{p^{st}_{i_{t-1}}(\lambda_{t})}
\end{equation}
\cite{PhysRevLett.104.090601}.
So if $\frac{p^{st}_{i_{t}}(\lambda_{t})}{p^{st}_{i_{t-1}}(\lambda_{t})}$ is replaced with $\frac{\epsilon_{i_{t}i_{t-1}}(\lambda_{t})}{\epsilon_{i_{t-1}i_{t}}(\lambda_{t})}$, the symmetric entropy production Eq. (\ref{symmetric_entropy}) is simply obtained.
Therefore nothing is different between them, but Eq. (\ref{symmetric_entropy}) is useful in some cases.

For an example, in the case of asymmetric random walk of ratio $w_{x+1,x}/w_{x,x+1}= \lambda$ in periodic 1-dimensional lattice space of size $L$, the symmetric and asymmetric operators can be set as
\begin{eqnarray}
\label{epsilon_D}
\epsilon_{x+1,x} &=& \epsilon_{x,x+1} = D \\
\label{nu_lambda}
\nu_{x+1,x} &=& \lambda, \quad \nu_{x,x+1} = 1,
\end{eqnarray}
where $D$ is diffusion constant.
Since the system will be evenly distributed in steady states of any $\lambda$, symmetric operator can be defined as Eq. (\ref{epsilon_D}).
Then time evolution operator can be set as $w_{x+1,x} = D \lambda$ and $w_{x,x+1} = D$, so asymmetric factor ca be defined as Eq. (\ref{nu_lambda}).
Because of Eq. (\ref{epsilon_D}), Eq. (\ref{symmetric_entropy}) becomes system entropy production, $\Delta S_{\epsilon} = \Delta S_{sys} = \ln p_{i_0}(0)/p_{i_T}(T)$.
So, after some algebra, symmetric entropy production rates are given by 
\begin{equation}
\langle \dot S_{\epsilon}(t) \rangle = \langle \dot S_{sys}(t) \rangle = - \int_0^{L} \dot p(x;t) \ln p(x;t) dx
\end{equation}
at time $t$.
As time passes, $\langle \dot S_{\epsilon} \rangle \rightarrow 0$ naturally because of diffusive motion, but only asymmetric part remains non-zero.
For fixed $\lambda$, from Eq. (\ref{asymmetric_entropy}) and (\ref{nu_lambda}), asymmetric path entropy production is given by 
\begin{equation}
\Delta S_{\nu} ([x]_0^T; \lambda) = T \ln \lambda
\end{equation}
since asymmetric entropy is generated at each time step over all space as much as $\ln \lambda$.
Dividing by $T$ after summing up over all possible paths, asymmetric entropy production rate is given by
\begin{equation}
\langle \dot S_{\nu} \rangle = D(\lambda-1) \ln \lambda
\end{equation}
for fixed $\lambda$.


\section{Conclusion and Discussions}

Before our works, Zia and Schmittmann have already decomposed flux in steady state into symmetric and asymmetric flux $linearly$ as $W_{ij} = S_{ij} + A_{ij}$, where $W_{ij} \equiv w_{ij}p^{st}_{j}$, $S_{ij} \equiv (W_{ij} + W_{ji})/2$ and $A_{ij} \equiv ( W_{ij} - W_{ji})/2$ are total, symmetric and asymmetric flux in steady states, respectively \cite{zia2006possible,zia2007probability}.
System and reservoir entropy production is obtained from $S_{ij}$ and $A_{ij}$, and their FTs are working well.


Although, we have tried a different way which is $log-decomposition$, because we believe that entropy production is to be decomposed properly by taking logarithm of dynamics since entropy is generally defined by logarithm of probability.
So we have tried log-decomposition of time evolution operator.
By the log-decomposition, not only path probability but also path entropy production is decomposed systematically as seen in TABLE \ref{tab:AS_separation}.


Symmetric operator $\epsilon_{ij}(\lambda)$ is given by a part satisfying detailed balance condition in steady states.
However, it must be remember that $\epsilon_{ij}(\lambda)$ does not mean actual equilibrium dynamics.
Symmetric operator can be defined only after $p^{st}_{i}(\lambda)$ is given.
That is a different point from equilibrium dynamics.
Symmetric operator can be defined from $p^{st}_{i}(\lambda)$ in the present work, whereas equilibrium distribution can be determined by Hamiltonian in equilibrium physics.
Even if given system had been in equilibrium, that can be driven out of equilibrium after the schedule of $\lambda$ begins.
Moreover generally $p^{st}_{i}(\lambda)$ is a nonequilibrium steady state.
Therefore $\epsilon_{ij}(\lambda)$ may be thought as effective equilibrium dynamics, but can not be actual equilibrium dynamics.


As mentioned above, entropy production is also decomposed into symmetric and asymmetric parts.
As seen in Eq.s (\ref{eq:S_epsilon}) and (\ref{eq:S_nu}), it can be said that $\Delta S_{na}$ is generated by symmetric (or reversible) dynamics, but $\Delta S_{a}$ is due to asymmetric (or irreversible) properties.
Therefore the relevant FTs of log-decomposition is definitely same with three detailed or integral FTs \cite{PhysRevLett.104.090601,PhysRevE.76.031132}, so there is no problem in log-decomposition.


\section*{Acknowledgement}

This works is supported by Basic Science Research Program through NRF grant 
funded by MEST (No. 2011-0014191).


\bibliographystyle{apsrev4-1}

\bibliography{coarse_sohn}

\end{document}